\documentstyle[amssymb,12pt,draft,epsf,palatino]{nature-pvd}

\def\simlt{\mathrel{\hbox{\rlap{\hbox{\lower4pt\hbox{$\sim$}}}\hbox{$<$}}}}
\def\simgt{\mathrel{\hbox{\rlap{\hbox{\lower4pt\hbox{$\sim$}}}\hbox{$>$}}}}

\def\ale{\mathrel{\hbox{\rlap{\hbox{\lower4pt\hbox{$\sim$}}}\hbox{$<$}}}}
\def\age{\mathrel{\hbox{\rlap{\hbox{\lower4pt\hbox{$\sim$}}}\hbox{$>$}}}}

\def\kms{km\,s$^{-1}$}
\def\msun{M$_{\odot}$}
\def\g1256{1255--0}

\def\spose#1{\hbox to 0pt{#1\hss}}
\newcommand\lsim{\mathrel{\spose{\lower 3pt\hbox{$\mathchar''218$}}
     \raise 2.0pt\hbox{$\mathchar''13C$}}}
\newcommand\gsim{\mathrel{\spose{\lower 3pt\hbox{$\mathchar''218$}}
     \raise 2.0pt\hbox{$\mathchar''13E$}}}

\begin{document}

\title{\LARGE \textsf{A substantial population of low mass stars
in\vspace{0.2cm}\\
luminous elliptical galaxies}}

\author{\textsf{Pieter G.\ van Dokkum}\affiliation[1]
  {\textsf{Astronomy Department, Yale University,  New Haven, CT, USA}}
   \textsf{\& Charlie Conroy}\affiliation[2]
   {\textsf{Department of Astrophysical Sciences, Princeton University, Princeton, NJ,
USA}}
\affiliation[3]
   {\textsf{Harvard-Smithsonian Center for Astrophysics, Cambridge, MA,
USA}}
%   {{\&} {Rachel Bezanson}}$^1$
   {}
}
\date{\today}{}
\headertitle{Low Mass Stars in Elliptical Galaxies}
\mainauthor{van Dokkum et al.}

\summary{The stellar initial mass function (IMF) describes the mass
  distribution of stars at the time of their formation and is of
  fundamental importance for many areas of astrophysics.  The IMF is
  reasonably well constrained in the disk of the Milky
  Way\cite{kroupa} but we have very little direct information
  on the form of the IMF in other galaxies and at earlier cosmic
  epochs.  Here we investigate the stellar mass function in elliptical
  galaxies by measuring the strength of the Na\,I
doublet\cite{faber:80,schiavon:97a} and the
  Wing-Ford molecular FeH band\cite{wf,schiavon:97}
in their spectra.  These lines
  are strong in stars with masses $\lesssim 0.3$\,\msun\ and weak or
  absent in all other types of stars.\cite{schiavon:97,schiavon:00,cushing:03}
  We unambiguously detect both
  signatures, consistent with previous studies\cite{couture:93}
that were based on data
  of lower signal-to-noise ratio. The direct detection of the light of
  low mass stars implies that they are very abundant in elliptical
  galaxies, making up $> 80$\,\% of the total number of stars and
  contributing $> 60$\,\% of the total stellar mass. We infer that the
  IMF in massive star-forming galaxies in the early Universe produced
  many more low mass stars than the IMF in the Milky Way disk, and was
  probably slightly steeper than the Salpeter form\cite{salp:55} in
  the mass range 0.1 -- 1\,\msun.}

\maketitle

%%%%%%%%%%%%%%%%%%%%%%%%%%%%%%%%%%%%%%%%%%%%%%%%%%%%%%%%%%%%%%%%%%%
%\onecolumn

We obtained spectra of eight of the most luminous and massive galaxies
in the nearby Universe: four of the brightest early-type
galaxies in the Virgo cluster and four in the Coma
cluster.  The galaxies were selected to have velocity dispersions
$\sigma>250$\,\kms, and were observed with the Low-Resolution Imaging
Spectrometer\cite{oke:95} (LRIS) on the Keck I telescope. In 2009 the
red arm of LRIS was outfitted with fully-depleted LBNL CCDs, which
have excellent sensitivity out to $\lambda>9000$\,\AA\ and virtually
no fringing.  The individual spectra of the four galaxies in each of
the two clusters were de-redshifted, averaged, and binned to a
resolution of 8\,\AA.

In Fig.\ 1b,c we show the spectral region near the $\lambda\lambda
8183,8195$ Na\,I doublet for the Virgo and Coma galaxies.  The doublet
appears as a single absorption feature due to Doppler broadening.  In
Fig.\ 1e,f we show the region around the $\lambda 9916$ Wing-Ford band
for the Virgo galaxies. This region could not be observed with
sufficient signal in Coma as it is redshifted to $1.015\,\mu$m. The
spectra are of very high quality.  The median $1\sigma$ scatter of the
four galaxies around the average spectrum is only $\sim 0.3$\,\% per
spectral bin. The median absolute difference between the Virgo and
Coma spectra is 0.4\,\% per spectral bin.

Both the Na\,I doublet and the Wing-Ford band are unambiguously
detected.  The central wavelength of the observed Na\,I line coincides
with the weighted average wavelength of the doublet and the observed
Wing-Ford band has the characteristic asymmetric profile reflecting
the $A\,^4\Delta - X\,^4\Delta$ transition of
FeH\cite{schiavon:97}. The Na\,I index is $0.058\pm 0.006$ mag in the
Virgo galaxies and $0.057\pm 0.007$ mag in Coma. The Wing-Ford index
in Virgo galaxies is $0.027\pm 0.005$.  The uncertainties are
determined from the scatter among the individual galaxies.  Note that
any residual systematic problems with the detector or atmosphere are
incorporated in this scatter, as the features were originally
redshifted to a different observed wavelength range for each of the
galaxies.

The immediate implication is that stars with masses $\lesssim
0.3$\,\msun\ are present in substantial numbers in the central regions
of elliptical galaxies. Such low mass stars are impossible to detect
individually in external galaxies, as they are too faint: Barnard's
star would have a $K$ band magnitude of 39 at the distance of the
Virgo cluster.  This in turn implies that there was a channel for
forming low mass stars in the progenitors of luminous early-type
galaxies in clusters. These star-forming progenitors are thought to be
relatively compact galaxies at $z=2-5$ with star formation rates of
10s or 100s of Solar masses per year. Some studies have suggested
truncated IMFs for such galaxies,\cite{baugh:05} with a cutoff below
1\,\msun. Such dwarfless IMFs are effectively ruled out by the
detection of the Na\,I line and the Wing-Ford band.

We turn to stellar population synthesis models to quantify the number
of low mass stars in elliptical galaxies. As discussed in detail in
the Supplementary Information, we use a flexible stellar population
synthesis code\cite{conroy} combined
with an extensive empirical library of stellar near-infrared
spectra.\cite{rayner:09} In Fig.\ 1 we show synthetic spectra in both
spectral regions for different choices of the
IMF,\cite{kroupa,salp:55,dokkum:08} including IMFs that are
steeper than the Salpeter form. The fits are
excellent, with differences between data and the best-fitting model
$\lesssim 0.5$\,\% over the entire spectral range. Predicted line
indices are compared to the observed values in Fig.\ 2.  The data
prefer IMFs with substantial dwarf populations. The best fits are
obtained for a logarithmic IMF slope of $x\sim -3$, a more dwarf-rich
(``bottom-heavy'') IMF than even the Salpeter form, which has
$x=-2.35$.  A Kroupa IMF (which is appropriate for the Milky Way) is
inconsistent with the Wing-Ford data at $> 2\sigma$ and inconsistent
with the Na\,I data at $> 4\sigma$, as are IMFs with even more
suppressed dwarf populations.\cite{dokkum:08,fardal:07,dave:08}
We note that the
$x=-3$ IMF also provides a much better fit to the region around
0.845$\mu$m than any of the other forms.  Taking the Salpeter IMF as a
limiting case, we find that stars with masses of 0.1\,\msun\ --
0.3\,\msun\ make up at least $80$\,\% of the total number of living
stars in elliptical galaxies, and contribute at least $60$\,\% of the
total stellar mass.

Although the formal uncertainty in the derived IMF slope is small we
stress that some unknown systematic effect could be present in the
stellar population synthesis modeling. In particular, weak features in
the spectra of giant stars in elliptical galaxies may be incorrectly
represented by the Milky Way giants that we use. It may also
be that the Na abundance of low mass stars in elliptical galaxies
is different from that of low mass mass stars in the Milky Way.
The fact that all
models fit the spectra of both the Virgo and Coma galaxies extremely
well outside of the IMF-sensitive regions gives some confidence in our
approach, and as we show in the Supplementary Information the quality of
the fit constrains possible contamination of particular elements such
as TiO.

Besides model uncertainties, the interpretation may
be complicated by the fact that we are
constraining the IMF some 10 billion years after the stars were
formed. 
It is now generally thought that elliptical galaxies have
undergone several (or many) mergers with other galaxies after their
initial collapse,\cite{naab:07} which may imply that the stellar
population is more complex than our single age, single metallicity
model. On the other hand, these
accretion events probably mostly add stars at large
radii\cite{dokkum:10} and may not have affected the core regions very
much. It will be interesting to search for gradients in
dwarf-sensitive features with more extensive data.\cite{boroson:91}

Our results are consistent with previous studies of the near-infrared
spectra of elliptical galaxies.\cite{couture:93,cenarro:03} They
are also consistent with recent dynamical and lensing
constraints on the IMF in elliptical galaxies with
large velocity dispersions\cite{treu:10}
and directly identify
the stars that are responsible for their high masses:
the dynamical data cannot distinguish dwarf-rich IMFs from
dwarf-deficient IMFs as the latter have a large amount of mass in
stellar remnants.\cite{dokkum:08} A steep IMF for elliptical galaxies is
also qualitatively consistent with the apparently higher number of
low mass stars in the Milky Way bulge than in the
disk.\cite{novati:08}
Our best-fitting IMF does not appear to be
consistent with the observed color and $M/L$ evolution of massive
cluster galaxies,\cite{dokkum:08} which suggest an IMF with a slope
$x\sim 0$ around $\sim 1$\,\msun.  Interpreting the evolution of the
colors and luminosities of elliptical galaxies relies on the
assumption that these galaxies evolve in a self-similar way, which may
not be valid.\cite{dokkum:10,vanderwel:08}
It could also be that the form of the
IMF is more complex than a powerlaw.

Our results also seem inconsistent with theoretical arguments for
dwarf-deficient IMFs at high redshift, which have centered on the idea
that the characteristic mass of stars scales with the Jeans mass in
molecular clouds\cite{larson:05,bate:05}.  The Jeans mass has a strong
temperature dependence and it has been argued that relatively high
ambient temperatures in high-redshift star forming galaxies may have
set a floor to the characteristic mass in the progenitors of
elliptical galaxies.\cite{larson:05,dokkum:08} However, the Jeans mass
also scales with density, and the gas densities in the star-forming
progenitors of the cores of elliptical galaxies were almost certainly
significantly higher than typical densities of star forming regions in
the Milky Way. Numerical simulations suggest that the formation of low
mass stars becomes inevitable if sufficiently high densities are
reached on sub-parsec scales.\cite{bonnell:08a} Furthermore, recent
semi-analyic models of the thermal evolution of gas
clouds
have emphasized the effects of dust-induced
cooling,\cite{schneider:10}
which is relatively insensitive to the ambient temperature
and particularly effective at high densities. Time-scale arguments
suggest that the physical conditions expected in starburst galaxies at
high redshift might even enhance low-mass star formation, rather than
suppress it.\cite{banerji:09}

Taken at face value, our results imply that the form of the IMF is not
universal but depends on the prevailing physical conditions:
Kroupa-like in quiet, star-forming disks and dwarf-rich in the
progenitors of massive elliptical galaxies. This informs models of
star formation and has important implications for the interpretation
of observations of galaxies in the early Universe. 
The stellar masses and star formation rates
of distant galaxies are usually estimated from
their luminosities, assuming some form of the
IMF.\cite{marchesini:09,labbe:10}
Our results suggest
that a different form should be used for different galaxies, greatly
complicating the analysis.
The bottom-heavy
IMF advocated here may also require a relatively low
fraction of dark matter within the central regions of nearby massive
galaxies.\cite{treu:10}

\bibliographystyle{nature-pap}
%\bibliography{journals,refs3}

\vspace{0.3cm}

\noindent
{\small \bf \textsf{Supplementary Information}}
\small{\textsf{is linked to the online version of the
paper at www.nature.com/nature.}}
\vspace{0.5cm}\\
{\small \bf {Acknowledgements}}
{\small 
  {We thank Rachel Bezanson, Jarle Brinchmann, Richard Larson,
  and Bob Zinn for
    discussions. We thank Ricardo Schiavon for insightful
comments which greatly improved
    the manuscript. This study is based on observations obtained at
    the W.~M.\ Keck Observatory. The authors wish to recognize and
    acknowledge the very significant cultural role and reverence that
    the summit of Mauna Kea has always had within the indigenous
    Hawaiian community. We are most fortunate to have the opportunity
    to conduct observations from this mountain.}}
\vspace{0.5cm}\\
{\small \bf {Author Contributions}}
{\small {P.v.D.\ obtained and analyzed the data and contributed to the
    analysis and interpretation.  C.C.\ constructed the stellar
    population synthesis models and contributed to the analysis and
    interpretation.}}
\vspace{0.5cm}\\
{\small \bf {Author Information}}
{\small {Reprints and permissions information is available at
npg.nature.com/reprintsandpermissions.
Correspondence and requests for materials
should be addressed to P.v.D.\ (pieter.vandokkum@yale.edu).}}

\clearpage

\noindent
{\small \bf {\textsf{Figure 1}} $|$ } 
{\small \bf {\textsf{Detection of the Na\,I doublet and the Wing-Ford band.}}}
{\small
{\bf a}, 
Spectra in the vicinity of the $\lambda\lambda{}8183,8195$ Na\,I doublet
for three stars from the IRTF library:$^{12}$ a K0 giant, which
dominates the light of old stellar populations; an M6 dwarf, whose
(small) contribution to the integrated light is sensitive to the form of
the IMF at low masses; and an M3 giant, which has potentially contaminating
TiO spectral features in this wavelength range.
{\bf b},
Averaged Keck/LRIS spectra of  NGC\,4261, NGC\,4374,
NGC\,4472, and NGC\,4649 in the Virgo cluster (black line) and NGC\,4840,
NGC\,4926, IC\,3976, and NGC\,4889 in the Coma cluster (grey line).
Four exposures of 180\,s were obtained for each galaxy.
The one-dimensional spectra were extracted from the reduced two-dimensional
data by summing the central $4''$, which
corresponds to $\approx 0.4$\,kpc at the distance of Virgo
and $\approx 1.8$\,kpc at the distance of Coma. We found little
or no dependence of the results on the choice of aperture.
Colored lines show stellar population synthesis models for
a dwarf-deficient ``bottom-light'' IMF\cite{dokkum:08}, a
dwarf-rich ``bottom-heavy'' IMF with $x=-3$, and an even more
dwarf-rich IMF. The models are for an age of
10\,Gyr and were smoothed to the average
velocity dispersion of the galaxies.
The $x=-3$ IMF fits the spectrum remarkably well.
{\bf c}, Spectra and models around the dwarf-sensitive Na\,I doublet.
A Kroupa IMF, which is appropriate for the Milky Way, does not
produce a sufficient number of low mass stars to explain the
strength of the absorption. An IMF steeper than Salpeter
appears to be needed.
{\bf d,e,f}, Spectra and models near the $\lambda9916$ Wing-Ford
band.  The observed Wing-Ford band also favors an IMF that is more
abundant in low mass stars than the Salpeter IMF.
All spectra and models were normalized by fitting low
order polynomials (excluding the feature of interest).
The polynomials were quadratic in {\bf a,b,d,e} and linear
in {\bf c,f}.
}
\vspace{0.5cm}

\noindent
{\small \bf {\textsf{Figure 2}} $|$ } 
{\small \bf {\textsf{Constraining the IMF.}}}
{\small
{\bf a}, 
Various stellar IMFs, ranging from a ``bottom-light'' IMF with
strongly suppressed dwarf formation\cite{dokkum:08} (light blue) to an
extremely ``bottom-heavy'' IMF with a slope $x=-3.5$. The IMFs are
normalized at 1\,\msun, as stars of approximately that mass dominate
the light of elliptical galaxies.  {\bf b}, Comparison of predicted
line Na\,I and Wing-Ford indices to the observed values. The indices
were defined analogous to refs.\ 4 and 8. The Na\,I index has central
wavelength $0.8195\,\mu$ and side bands at $0.816\,\mu$m and
$0.825\,\mu$m. The Wing-Ford index has central wavelength
$0.992\,\mu$m and side bands at $0.985\,\mu$m and $0.998\,\mu$m. The
central bands and side bands are all 20\,\AA\ wide. Both observed line
indices are much stronger than expected for a Kroupa IMF. The best
fits are obtained for IMFs that are slightly steeper than Salpeter.  }

\newpage
\begin{figure}
\epsfxsize=16cm
\epsffile{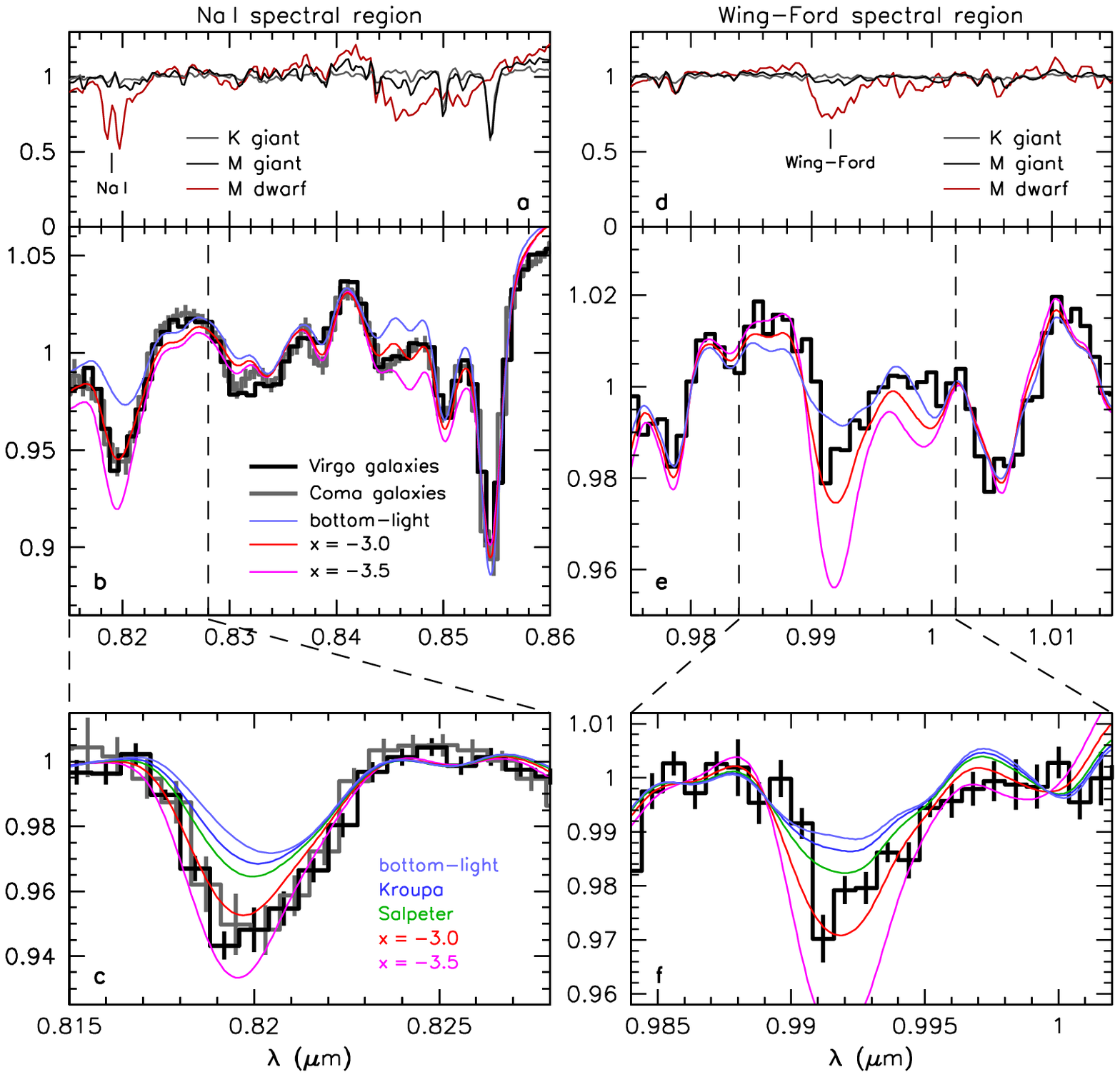}
\end{figure}

%\newpage
\begin{figure}
\epsfxsize=16cm
\epsffile{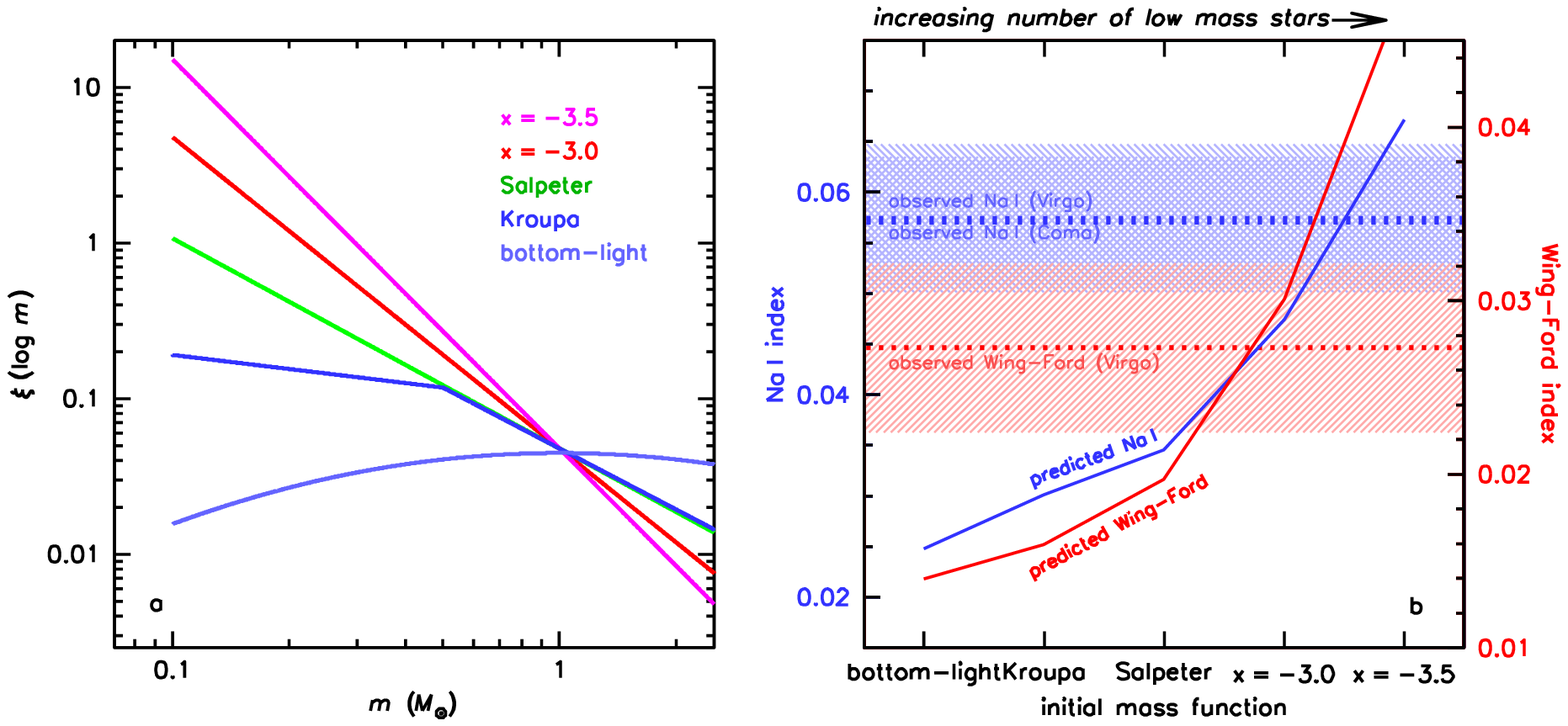}
\end{figure}

\newpage

\onecolumn
\noindent
{\bf Supplementary Information: Stellar population synthesis modeling}
\vspace{0.2cm}\\
The stellar population model used in the present analysis was
constructed specifically for this project.  We therefore describe
it in some detail below.
\vspace{0.2cm}\\
{\em Construction of the model:}
\vspace{0.1cm}\\
The IRTF spectral library\cite{cushing:05,rayner2:09} constitutes the
core of our model.  This spectral library is novel in its near-IR
coverage of a large sample of cool stars at a resolving power of
$R\sim 2000$.  65 stars were selected from this library with
luminosity classes III and V (giants and dwarfs).  Peculiar stars were
removed, including stars with non-solar Fe abundance and/or other
peculiar abundance patterns (mainly Ca, CN, CH, and Ba), as well as
known pulsating M giants.  Stars with significant reddening were also
removed.  The IRTF library includes 2MASS $JHK$ and literature $V$
band photometry for each star.

We estimated effective temperatures for these stars based on empirical
$V-K$ vs.\ $T_{\rm eff}$ relations that have been separately
determined for giants\cite{alonso:99,perrin:98,ridgeway:80}
and dwarfs.\cite{alonso:96,leggett:96}
The adopted dwarf relations
agree to within 100\,K with a recently updated temperature
scale.\cite{casagrande:08,casagrande:10}

The IRTF library is absolute-flux calibrated against 2MASS photometry.
We adopt parallax values obtained from SIMBAD in order to derive
absolute luminosities.  In order to derive bolometric luminosities,
which we will use to assign stellar masses, we must extrapolate the
IRTF library beyond the observed range.  We perform the extrapolation
by utilizing the latest version of the PHOENIX model spectra.  The
model spectra are normalized to the IRTF library at the blue and red
ends of the observed wavelength range so that the extrapolation is
continuous across the transition region.  The model spectrum used to
extrapolate each IRTF stellar spectrum was chosen from the estimated
$T_{\rm eff}$ for each IRTF star and an estimate of the surface
gravity of the star based on its luminosity class.  We note that the
extrapolation does not lead to large uncertainties as the IRTF
wavelength range ($0.8-2.5\,\mu$m) encompasses the majority of the
bolometric flux, especially for K and M stars. As an example, 80\,\%
of the flux from M-type stars is emitted within the IRTF spectral
range.

The resulting $T_{\rm eff}$ and $L_{\rm bol}$ for each IRTF star is
plotted in Supp.\ Fig.\ 1.  Also shown in this Figure are model
isochrones (described below).  The agreement between the data and
models is remarkable, and provides a valuable consistency check on
both the estimated $T_{\rm eff}$ and $L_{\rm bol}$ values.

We adopt a combination of isochrones in order to utilize the most
accurate stellar models for each range in stellar mass.  At the lowest
masses we adopt the Baraffe  stellar evolution calculations,\cite{baraffe:98}
which are unique in their use of realistic atmospheric boundary
conditions.  Between 0.2\,\msun\ and the tip of the RGB we use the
Dartmouth evolutionary calculations\cite{dartmouth}, which have been
shown to produce accurate fits to photometry of open and globular
clusters of a variety of ages and metallicities.\cite{an:09}
The Padova evolutionary tracks\cite{marigo:08} are used to
extend the isochrones through the horizontal branch and asymptoptic
giant branch evolutionary phases.  This combined isochrone set is
shown in Supp.\ Fig.\ 1.

The isochrones are used to assign masses to each IRTF spectrum, based
on the measured $L_{\rm bol}$ for each star.  We note that the stellar
$T_{\rm eff}$ values are used only to choose the appropriate PHOENIX
model spectra to extrapolate the observed spectra.  Our conclusions
remain unchanged if we instead assign masses based on the estimated
$T_{\rm eff}$ for each star.  Alpha-enhancements of 0.2 dex have a
very minor effect on the $L_{\rm bol}-M$ and $T_{\rm eff}-M$
relations, when [Fe/H] is held constant.  As a final test of our
procedure we compare the masses to those predicted from
empirical
relations between mass and $K-$band luminosity derived for
stars in binary systems.\cite{henry:93,delfosse:00}  Such relations are
completely independent of stellar evolution calculations because the
masses are derived from dynamical modeling.  The stellar mass --
$K-$band luminosity relation for our IRTF stars agrees very well with
the relations based on dynamical masses in the range
$0.08<M<0.6\,$\msun.  The agreement is particularly strong when
bolometric luminosities are used to assign masses, which motivates the
use of bolometric luminosities as our default method.
Based on these comparisons, we conclude that our mass
estimates are reliable.

Once masses are assigned, a
synthesized spectrum is created by integrating the observed spectra
over an initial mass function.
\vspace{0.2cm}\\
{\em Metallicity and $\alpha$-enhancement:}
\vspace{0.1cm}\\
The empirical spectral library that we use has important advantages
over libraries of model atmospheres. Model atmospheres do not
reproduce molecular features such as TiO very well, which makes it
difficult to interpret fits of these models to our spectra. We note
here that Kurucz models --- which are used in most existing stellar
population synthesis models at $\lambda>8000$\,\AA\ --- also prefer
IMFs that are more ``bottom-heavy'' than the Kroupa IMF, but give much
poorer fits than the IRTF empirical library to the spectral regions
away from Na\,I and the Wing-Ford band.

A disadvantage of the empirical library is that it is not
straightforward to assess the effects of abundance variations.  We
know that the iron abundance in the eight Coma and Virgo ellipticals
is approximately Solar\cite{trager:00,thomas:05},
and so the value of [Fe/H] of the Milky Way
stars is appropriate for the galaxies that we study here. However, the
abundance of $\alpha$-elements (and the overall mass-weighted
metallicity [Z/H]) is higher in massive elliptical galaxies than in
the Milky Way stars.  Although sodium (responsible for the Na\,I line)
and iron (responsible for the Wing-Ford band) are themselves not
$\alpha$-elements, both lines and their sidebands overlap with TiO
lines that are present in M giants. 

We empirically determined to what extent enhanced TiO absorption could
influence the results by creating models with artifically enhanced
late M giant light.  The model shown in Supp.\ Fig.\ 2 has a Kroupa
IMF and an additional contribution of late M giants amounting to
40\,\% of the light at 0.9$\,\mu$m. This is an extreme model as these
stars should contribute only a few percent to the light, but we
use it here to assess the effects of enhancing TiO features. This model
fits the observed strength of the Wing-Ford band quite well as it
coincides with the $\delta(2-3)$ band of TiO. However, it also
predicts that other TiO features are nearly as strong as
the absorption at $0.99\,\mu$m, which is clearly not the case.
Moreover, the fit to the Na\,I line is not improved as the only
relevant TiO feature is slightly
redward of Na\,I, and the fit to the rest of
the spectrum in the wavelength range $0.815-0.860\,\mu$m is
unacceptable. We note that the apparently limited enhancement of
TiO with respect to the Milky Way giants is consistent with
the weak response of TiO to either metallicity or $\alpha$-enhancement
in the PHOENIX library and the models of ref.\ 18.

We infer that the Wing-Ford band should be interpreted
with caution but that the combination of the Wing-Ford band with 
TiO features at other wavelengths
and (particularly) the Na\,I line should be a
robust indicator of the presence of low mass stars. This is fully
consistent with previous theoretical work.\cite{schiavon2:97a}

\newpage

\noindent
{\small \bf {\textsf{Supplementary Figure 1}} $|$ } 
{\small \bf {\textsf{The IRTF stellar library.}}}
{\small
Hertzprung-Russell diagram of IRTF stars and theoretical stellar
evolution calculations.  IRTF stellar $L_{\rm bol}$ and $T_{\rm
  eff}$ have been estimated as described in the text.  Theoretical
models assume [Fe/H]\,$=0.0$ and an age of 13.7 Gyr.  The models
encompass all phases of stellar evolution from main sequence hydrogen
burning through the end of the asymptotic giant branch, and extend
down to the hydrogen burning limit of 0.08\,\msun.
}
\vspace{0.5cm}

\noindent
{\small \bf {\textsf{Supplementary Figure 2}} $|$ } 
{\small \bf {\textsf{Effect of enhancing M giant features.}}}
{\small
{\bf a} 
Spectra and models around the Na\,I doublet.  Our favored model with a
steep IMF ($x=-3.0$) is compared to a model that assumes a Kroupa IMF
with the addition of a 40\% light contribution from an M6 giant
star.  Since late M giants have very strong TiO features this
exercise is meant to mimic the effects of increasing
$\alpha$-enhancement in the models.  Clearly the addition of a
substantial amount of M giant light provides a poor fit to
the absorption line at $\approx 0.82\,\mu$m as the central
wavelength is not matched. This demonstrates that
this feature is due to Na\,I and not TiO. The fit is also very poor
to the rest of the spectrum in this wavelength range.
{\bf b}
Spectra and models around the Wing-Ford band.  Here the addition of M
giant light adequately reproduces the strong observed Wing-Ford
absorption due to the TiO absorption feature that coincides with the
FeH absorption.  However, this fit also predicts unacceptably large
absorption at $0.983\,\mu$m, $1.001\,\mu$m, and $1.006\,\mu$m as
these TiO features are nearly as strong as
the one at $0.992\,\mu$m.
}
\vspace{0.5cm}

\newpage

\begin{figure}
\epsfxsize=12cm
\epsffile{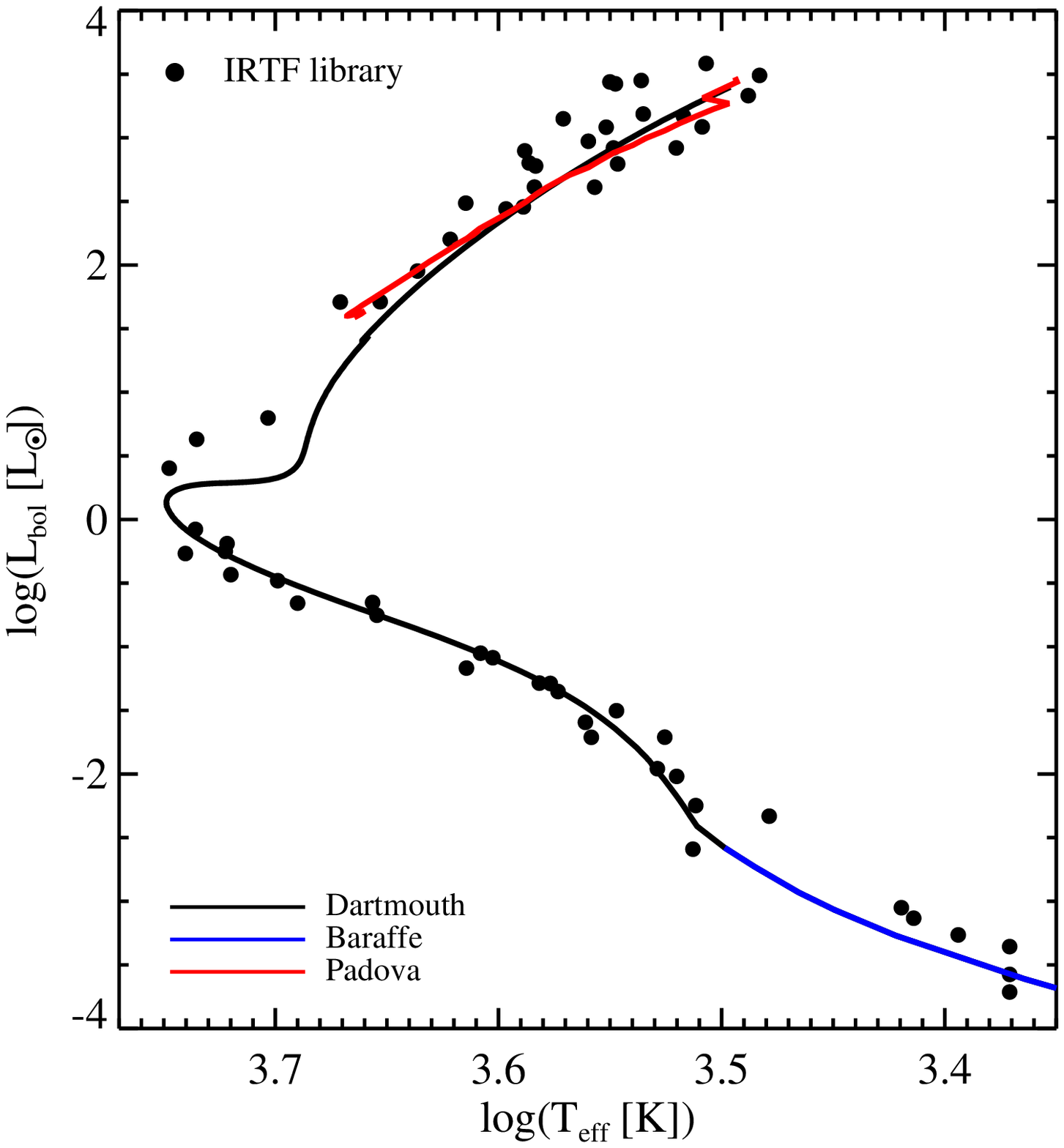}
\end{figure}

\begin{figure}
\epsfxsize=16cm
\epsffile{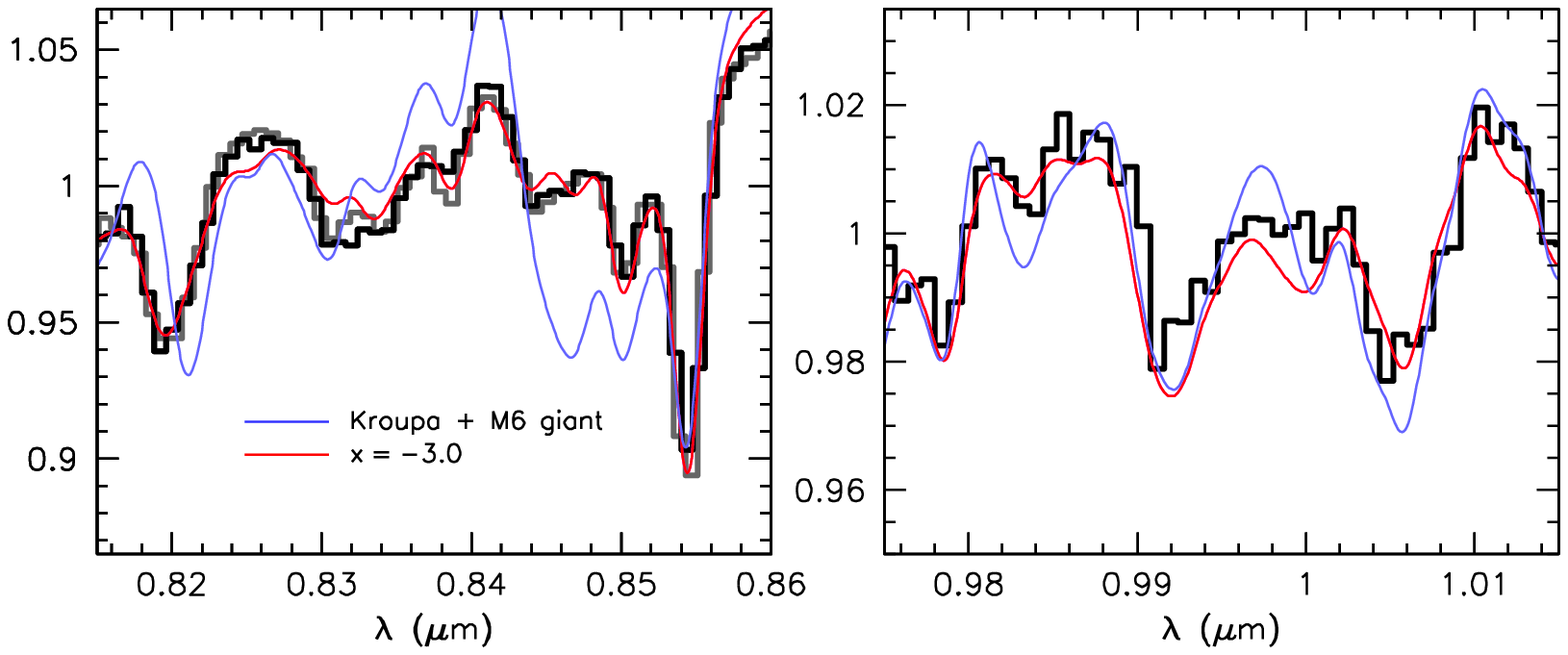}
\end{figure}


\begin{thebibliography}{10}

\bibitem[{Kroupa}<1>]{kroupa}
Kroupa, P.
{On the variation of the initial mass function.}
\newblock {\it Mon.\ Not.\ R.\ Astron.\ Soc.} {\bf 322}, 231--246 (2001)

\bibitem[{Faber \& French}<2>]{faber:80}
Faber, S.~M., \& French, H.~B.
{Possible M dwarf enrichment in the semistellar nucleus of M31.}
\newblock {\it Astrophys.\ J.}, {\bf 235}, 405--412 (1980)


\bibitem[{Schiavon et al.}<3>]{schiavon:97a}
Schiavon, R.~P., Barbuy, B., Rossi, S.~C.~F., \&
Milone, A.
{The Near-Infrared Na\,I Doublet Feature in M Stars.}
\newblock {\it Astrophys.\ J.} {\bf 479}, 902--908 (1997)

\bibitem[{Wing \& Ford}<4>]{wf}
Wing, R.~F., \& Ford, W.~K.
{The Infrared Spectrum of the Cool Dwarf Wolf 359.}
\newblock {\it Publ.\ Astron.\ Soc.\ Pacific} {\bf 81}, 527--529 (1969)


\bibitem[{Schiavon}{\it et~al.}<5>]{schiavon:97}
Schiavon, R.~P., Barbuy, B., \& Singh, P.~D.
{The FeH Wing-Ford Band in Spectra of M Stars.}
\newblock {\it Astrophys.\ J.} {\bf 484}, 499--510 (1997)

\bibitem[{Schiavon}{\it et~al.}<6>]{schiavon:00}
Schiavon, R.~P., Barbuy, B., \& Bruzual, A.~G.
{Near-Infrared Spectral Features in Single-aged Stellar
Populations.}
\newblock {\it Astrophys.\ J.} {\bf 532}, 453--460 (2000)

\bibitem[{Cushing}{\it et~al.}<7>]{cushing:03}
Cushing, M.~C., Rayner, J.~T., Davis, S.~P., \& Vacca, W.~D.
{FeH Absorption in the Near-Infrared Spectra of Late M and L Dwarfs.}
\newblock {\it Astrophys.\ J.} {\bf 582}, 1066--1072 (2003)

\bibitem[{Couture \& Hardy}<8>]{couture:93}
Couture, J., \& Hardy, E.
{The low-mass stellar content of galaxies -- Constraints through
hybrid population synthesis near 1 micron.}
\newblock {\it Astrophys.\ J.} {\bf 406}, 142--157 (1993)

\bibitem[{Salpeter}<9>]{salp:55}
Salpeter, E.~E.
{The Luminosity Function and Stellar Evolution.}
\newblock {\it Astrophys.\ J.} {\bf 121}, 161--167 (1955)

\bibitem[{Oke} {\it et~al.}<10>]{oke:95}
Oke, J.~B., Cohen, J.~G., Carr, M., Cromer, J., Dingizian, A.
{\it et al.}
{The Keck Low-Resolution Imaging Spectrometer.}
\newblock {\it Publ.\ Astron.\ Soc.\ Pacific} {\bf 107}, 375--385 (1995)

\bibitem[{Baugh} {\it et~al.}<11>]{baugh:05}
Baugh, C.~M., Lacey, C.~G., Frenk, C.~S., Granato, G.~L., Silva, L.
{\it et al.}
{Can the faint submillimetre galaxies be explained in the $\Lambda$
cold dark metter model?}
\newblock {\it Mon.\ Not.\ R.\ Astron.\ Soc.} {\bf 356}, 1191--1200 (2005)

\bibitem[{Conroy}{\it et~al.}<12>]{conroy}
Conroy, C., Gunn, J.~E., \& White, M.
{The Propagation of Uncertainties in Stellar Population Synthesis Modeling. I. The Relevance of Uncertain Aspects of Stellar Evolution and the Initial Mass Function to the Derived Physical Properties of Galaxies.}
\newblock {\it Astrophys.\ J.} {\bf 699}, 486--506 (2009)

\bibitem[{Rayner et al.}<13>]{rayner:09}
Rayner, J.~T. and Cushing, M.~C. and Vacca, W.~D.
{The Infrared Telescope Facility (IRTF) Spectral Library: Cool Stars.}
\newblock {\it Astrophys.\ J.} {\bf 623}, 289--432 (2009)


\bibitem[{van Dokkum}<14>]{dokkum:08}
van Dokkum, P.~G.
{Evidence of Cosmic Evolution of the Stellar Initial Mass Function.}
\newblock {\it Astrophys.\ J.} {\bf 674}, 29--50 (2008)

\bibitem[{Fardal} {\it et~al.}<15>]{fardal:07}
Fardal, M.~A., Katz, N., Weinberg, D.~H., \& Dav\'e, R.
{On the evolutionary history of stars and their fossil mass and light.}
\newblock {\it Mon.\ Not.\ R.\ Astron.\ Soc.} {\bf 379}, 985--1002 (2007)

\bibitem[{Dave}<16>]{dave:08}
Dav\'e, R.
{The galaxy stellar mass-star formation rate relation: evidence
for an evolving stellar initial mass function?}
\newblock {\it Mon.\ Not.\ R.\ Astron.\ Soc.} {\bf 385}, 147--160 (2008)

\bibitem[{Naab et al.}<17>]{naab:07}
Naab, T., Johansson, P.~H., Ostriker, J.~P., \& Efstathiou, G.
{Formation of Early-Type Galaxies from Cosmological Initial
Conditions.}
\newblock {\it Astrophys.\ J.} {\bf 658}, 710--720 (2007)

\bibitem[{van Dokkum et al.}<18>]{dokkum:10}
van Dokkum, P.~G., Whitaker, K.~E., Brammer, G., Franx, M.,
Kriek, M., {\it et al.}
{The Growth of Massive Galaxies Since $z=2$.}
\newblock {\it Astrophys.\ J.} {\bf 709}, 1018--1041 (2010)

\bibitem[{Boroson \& Thompson}<19>]{boroson:91}
Boroson, T.~A., \& Thompson, I.~B.
{Color distributions in early type galaxies. III -- Radial
gradients in spectral features.}
\newblock {\it Astron.\ J.} {\bf 101}, 111--126 (1991)

\bibitem[{Cenarro et al.}<20>]{cenarro:03}
Cenarro, A.~J., Gorgas, J., Vazdekis, A., Cardiel, N.,
\& Peletier, R.~F. {Near-infrared line-strengths in elliptical
galaxies: evidence for initial mass function variations?}
\newblock {\it Mon.\ Not.\ R.\ Astron.\ Soc.} {\bf 339}, L12--L16 (2003)

\bibitem[{Treu et al.}<21>]{treu:10}
Treu, T., Auger, M.~W., Koopmans, L.~V.~E., Gavazzi, R., Marshall,
P.~J., \& Bolton, A.~S.
{The Initial Mass Function of Early-Type Galaxies.}
\newblock {\it Astrophys.\ J.} {\bf 709}, 1195--1202 (2010)

%\bibitem[{Zoccali et al.}<20>]{zoccali:00}
%Zoccali, M., Cassisi, S., Frogel, J.~A., Gould, A., Ortolani, S.,
%Renzini, A., Rich, R.~M., \& Stephens, A.~W.
%{The Initial Mass Function of the Galactic Bulge down to $\sim
%0.15\,$M$_{\odot}$.}
%\newblock {\it Astrophys.\ J.} {\bf 530}, 418--428 (2000)

\bibitem[{Calchi Novati et al.}<22>]{novati:08}
Calchi Novati, S., de Luca, F., Jetzer, Ph., Mancini, L., \&
Scarpetta, G.
{Microlensing constraints on the Galactic bulge initial mass function.}
\newblock {\it Astron.\ Astrophys.}, {\bf 480}, 723--733

\bibitem[{van der Wel et al.}<23>]{vanderwel:08}
van der Wel, A., Holden, B.~P., Zirm, A.~W., Franx, M., Rettura, A.,
Illingworth, G.~D., \& Ford, H.~C.
{Recent Structural Evolution of Early-Type Galaxies: Size Growth
from $z=1$ to $z=0$.}
\newblock {\it Astrophys.\ J.}, {\bf 688}, 48--58

\bibitem[{Larson}<24>]{larson:05}
Larson, R.~B.
{Thermal physics, coud geometry and the stellar initial mass function.}
\newblock {\it Mon.\ Not.\ R.\ Astron.\ Soc.} {\bf 359}, 211--222 (2005)

\bibitem[{Bate \& Bonnell}<25>]{bate:05}
Bate, M.~R., \& Bonnell, I.~A.
{The origin of the initial mass function and its dependence on the mean
Jeans mass in molecular clouds.}
\newblock {\it Mon.\ Not.\ R.\ Astron.\ Soc.} {\bf 356}, 1201--1221 (2005)

\bibitem[{Bonnell et al.}<26>]{bonnell:08a}
Bonnell, I.~A., Clark, P., \& Bate, M.~R.
{Gravitational fragmentation and the formation of brown dwarfs in stellar
clusters.}
\newblock {\it Mon.\ Not.\ R.\ Astron.\ Soc.} {\bf 389}, 1556--1562 (2008)

\bibitem[{Schneider \& Omukai}<27>]{schneider:10}
Schneider, R., \& Omukai, K.
{Metals, dust and the cosmic microwave background: fragmentation
of high-redshift star-forming clouds.}
\newblock {\it Mon.\ Not.\ R.\ Astron.\ Soc.} {\bf 402}, 429--435 (2010)

\bibitem[{Banerji et al.}<28>]{banerji:09}
Banerji, S., Viti, S., Williams, D.~A., \& Rawlings, J.~M.~C.
{Timescales for Low-Mass Star Formation in Extragalactic Environments: Implications
for the Stellar Initial Mass Function.}
\newblock {\it Astrophys.\ J.} {\bf 692}, 283--289 (2009)

\bibitem[{Marchesini et al.}<29>]{marchesini:09}
Marchesini, D., van Dokkum, P.~G., F\"orster Schreiber, N.~M.,
Franx, M., Labb\'e, I., \& Wuyts, S.
{The Evolution of the Stellar Mass Function of Galaxies from $z=4.0$
and the First Comprehensive Analysis of its Uncertainties: Evidence
for Mass-Dependent Evolution.}
\newblock {\it Astrophys.\ J.} {\bf 701}, 1765--1796 (2009)

\bibitem[{Labb\'e et al.}<30>]{labbe:10}
Labb\'e, I., Gonz\'alez, V., Bouwens, R.~J., Illingworth, G.~D.,
Franx, M., Trenti, M., Oesch, P.~A., et al.
{Star Formation Rates and Stellar Masses of $z=7–8$
Galaxies from IRAC Observations of the WFC3/IR Early
Release Science and the HUDF Fields.}
\newblock {\it Astrophys.\ J.} {\bf 716}, L103--L108 (2010)


%\bibitem[{Hardy \& Couture}<5>]{hardy:88}
%Hardy, E., \& Couture, J.
%{Detection and measurement of the Wing-Ford band in the
%near-infrared spectra of elliptical galaxies.}
%\newblock {\it Astrophys.\ J.} {\bf 325}, L29--L31



%\bibitem[{Worthey} {\it et~al.}<3>]{worthey:92}
%Worthey, G., Faber, S.~M., \& Gonzalez, J.~J.
%{Mg and Fe absorption features in elliptical galaxies.}
%\newblock {\it Astroph.\ Journal} {\bf 398}, 69--73 (1992)


%\bibitem[{Tumlinson}<6>]{tumlinson:07}
%Tumlinson, J.
%{Carbon-Enhanced Metal-poor Stars, the Cosmic Microwave Background, and the Stellar Initial Mass Function in the Early Universe.}
%\newblock {\it Astroph.\ Journal} {\bf 664}, L63--L66 (2007)




%\bibitem[{Whitford}<12>]{whitford:77}
%Whitford, A.~E.
%{The Wing-Ford band as a constraint on the mass function in old
%galaxy populations.}
%\newblock {\it Astroph.\ Journal} {\bf 211}, 527--538 (1977)




%\bibitem[{Frogel}{\it et~al.}<15>]{frogel:75}
%Frogel, J.~A., Becklin, E.~E., Neugebauer, G., Matthews, K., Persson,
%S.~E., \& Aaronson, M.
%{Stellar content of the nuclei of elliptical galaxies determined
%from the 2.3-micron CO band strengths.}
%\newblock {\it Astroph.\ Journal} {\bf 195}, L15--L18


%\bibitem[{van der Marel \& van Dokkum}<16>]{vdmvd:07}
%van der Marel, R.~P., \& van Dokkum, P.~G.
%{Dynamical Models of Elliptical Galaxies in $z=0.5$ Clusters.
%II. Mass-to-Light Ratio Evolution without Fundamental Plane
%Assumptions.}
%\newblock {\it Astroph.\ Journal} {\bf 668}, 756--771 (2007)



%\bibitem[{Bruzual \& Charlot}<19>]{bc03}
%Bruzual, G., \& Charlot, S.
%{Stellar population synthesis at the resolution of 2003.}
%\newblock {\it Mon.\ Not.\ Royal Ast.\ Soc.} {\bf 344}, 1000--1028 (2003)

%\bibitem[{Maraston}<20>]{m05}
%Maraston, C.
%{Evolutionary population synthesis: models, analysis of the ingredients
%and application to high-z galaxies.}
%\newblock {\it Mon.\ Not.\ Royal Ast.\ Soc.} {\bf 362}, 799--825 (2005)

%\bibitem[{Leggett et al.}<21>]{leggett:00}
%Leggett, S.~K., Allard, F., Dahn, C., Hauschildt, P.~H., Kerr, T.~H.,
%\& Rayner, J.
%{Spectral Energy Distributions for Disk and Halo M Dwarfs.}
%\newblock {\it Astroph.\ Journal} {\bf 535}, 965--974 (2000)

%\bibitem[{Pickles}<23>]{pickles}
%Pickles, A.~J.
%{A Stellar Spectral Flux Library: 1150--25000\,\AA}
%\newblock {\it Publ.\ Astr.\ Soc.\ Pac.} {\bf 110}, 863--878 (1998)

%\bibitem[{Trager et al.}<24>]{trager:00}
%Trager, S.~C., Faber, S.~M., Worthey, G., \& Gonz\'alez, J.~J.
%{The Stellar Population Histories of Local Early-Type Galaxies. I. Population
%Parameters.}
%\newblock {\it Astron.\ Journal} {\bf 119}, 1645--1676 (2000)








%\bibitem[{Bonnell \& Rice}<29>]{bonnell:08b}
%Bonnell, I.~A., \& Rice, W.~K.~M.
%{Star Formation Around Supermassive Black Holes.}
%\newblock {\it Science} {\bf 321}, 1060--1062 (2008)

\end{thebibliography}

\begin{thebibliography}{10}


\bibitem[{Cushing et al.}<1>]{cushing:05}
Cushing, M.~C., Rayner, J.~T., \& Vacca, W.~D.
{An Infrared Spectroscopic Sequence of M, L, and T Dwarfs.}
\newblock {\it Astrophys.\ J.} {\bf 623}, 1115--1140 (2005)

\bibitem[{Rayner et al.}<2>]{rayner2:09}
Rayner, J.~T., Cushing, M.~C., \& Vacca, W.~D.
{The Infrared Telescope Facility (IRTF) Spectral Library: Cool Stars.}
\newblock {\it Astrophys.\ J.} {\bf 623}, 289--432 (2009)

\bibitem[{Alonso et al.}<3>]{alonso:99}
Alonso, A., Arribas, S., \& Mart{\'{\i}}nez-Roger, C.
{The effective temperature scale of giant stars (F0-K5). II. Empirical calibration of $T_{\rm eff}$
versus colours and [Fe/H].}
\newblock {\it Astron.\ Astrophys.} {\bf 140}, 261--277 (1999)

\bibitem[{Perrin et al.}<4>]{perrin:98}
Perrin, G., Coud{\'e} du Foresto, V., Ridgway, S.~T.,
	Mariotti, {J.-M.}, Traub, W.~A., Carleton, N.~P., \&
	Lacasse, M.~G.
{Extension of the effective temperature scale of giants to types later than M6.}
\newblock {\it Astron.\ Astrophys.} {\bf 331}, 619--626 (1998)

\bibitem[{Ridgway et al.}<5>]{ridgeway:80}
Ridgway, S.~T., Joyce, R.~R., White, N.~M., \& Wing, R.~F.
{Effective temperatures of late-type stars - The field giants from K0 to M6.}
\newblock {\it Astrophys.\ J.} {\bf 235}, 126--137 (1980)

\bibitem[{Alonso et al.}<6>]{alonso:96}
Alonso, A., Arribas, S., \& Martinez-Roger, C.
{The empirical scale of temperatures of the low main sequence (F0V-K5V).}
\newblock {\it Astron.\ Astrophys.} {\bf 313}, 873--890 (1996)

\bibitem[{Leggett et al.}<7>]{leggett:96}
Leggett, S.~K., Allard, F., Berriman, G., Dahn, C.~C.,\&
	Hauschildt, P.~H.
{Infrared Spectra of Low-Mass Stars: Toward a Temperature Scale for Red Dwarfs.}
\newblock {\it Astrophys.\ J.} {\bf 104}, 117 (1996)

\bibitem[{Casagrande et al.}<8>]{casagrande:08}
Casagrande, L., Flynn, C., \& Bessell, M.
{M dwarfs: effective temperatures, radii and metallicities.}
\newblock {\it Mon.\ Not.\ R.\ Astron.\ Soc.} {\bf 389}, 585--607 (2008)

\bibitem[{Casagrande et al.}<9>]{casagrande:10}
Casagrande, L., Ram{\'{\i}}rez, I., Mel{\'e}ndez, J.,
	Bessell, M., \& Asplund, K.
{An absolutely calibrated $T_{\rm eff}$ scale from the infrared flux method. Dwarfs and subgiants.}
\newblock {\it Mon.\ Not.\ R.\ Astron.\ Soc.} {\bf 512}, 54 (2010)

\bibitem[{Baraffe et al.}<10>]{baraffe:98}
Baraffe, I., Chabrier, G., Allard, F., \& Hauschildt, P.~H.
{Evolutionary models for solar metallicity low-mass stars: mass-magnitude relationships and color-magnitude diagrams.}
\newblock {\it Astron.\ Astrophys.} {\bf 337}, 403--412 (1998)

\bibitem[{Dotter et al.}<11>]{dartmouth}
Dotter, A., Chaboyer, B., Jevremovi\'c, D., Kostov, V., Baron,
E., \& Ferguson, J.~W.
{The Dartmouth Stellar Evolution Database.}
\newblock {\it Astrophys.\ J.\ Supp.} {\bf 178}, 89--101 (2008)

\bibitem[{An et al.}<12>]{an:09}
An, D., Pinsonneault, M.~H., Masseron, T., Delahaye, F.,
	Johnson, J.~A., Terndrup, D.~M., Beers, T.~C.,
	Ivans, I.~I., \& Ivezi{\'c}, {\v Z}
{Galactic Globular and Open Clusters in the Sloan Digital Sky Survey. II. Test of Theoretical Stellar Isochrones.}
\newblock {\it Astrophys.\ J.} {\bf 623}, 523--544 (2009)

\bibitem[{Marigo et al.}<13>]{marigo:08}
Marigo, P., Girardi, L., Bressan, A., Groenewegen, M.~A.~T.,
	Silva, L., \& Granato, G.~L.
{Evolution of asymptotic giant branch stars. II. Optical to far-infrared isochrones with improved TP-AGB models.}
\newblock {\it Astron. Astrophys.} {\bf 482}, 883--905 (2008)

\bibitem[{Henry \& McCarthy}<14>]{henry:93}
Henry, T.~J. \& McCarthy, Jr., D.~W.
{The mass-luminosity relation for stars of mass 1.0 to 0.08 solar mass.}
\newblock {\it Astron.\ J.} {\bf 106}, 773--789 (1993)

\bibitem[{Delfosse et al.}<15>]{delfosse:00}
Delfosse, X., Forveille, T., S{\'e}gransan, D.,
  Beuzit, J.-L., Udry, S., Perrier, C., \& Mayor, M.
{Accurate masses of very low mass stars. IV. Improved mass-luminosity
relations.}
\newblock {\it Astron.\ Astrophys.} {\bf 364}, 217--224 (2000)

\bibitem[{Trager et al.}<16>]{trager:00}
Trager, S.~C., Faber, S.~M., Worthey, G., \& Gonz\'alez, J.~J.
{The Stellar Population Histories of Local Early-Type Galaxies.
I. Population Parameters.}
\newblock {\it Astrophys.\ J.} {\bf 119}, 1645--1676 (2000)

\bibitem[{Thomas et al.}<17>]{thomas:05}
Thomas, D., Maraston, C., Bender, R., \& Mendes de Oliveira, C.
{The Epochs of Early-Type Galaxy Formation as a Function of Environment.}
\newblock {\it Astrophys.\ J.} {\bf 621}, 673--694 (2005)

\bibitem[{Thomas et al.}<18>]{thomas:03}
Thomas, D., Maraston, C., \& Bender, R.
{Stellar Population models of Lick indices with variable element
abundance ratios.}
\newblock {\it Mon.\ Not.\ R.\ Astron.\ Soc.} {\bf 339}, 897--911 (2003)

\bibitem[{Schiavon et al.}<19>]{schiavon2:97a}
Schiavon, R.~P., Barbuy, B., Rossi, S.~C.~F., \&
Milone, A.
{The Near-Infrared Na\,I Doublet Feature in M Stars.}
\newblock {\it Astrophys.\ J.} {\bf 479}, 902--908 (1997)

%\bibitem[{Kroupa}<1>]{kroupa}
%Kroupa, P.
%{On the variation of the initial mass function.}
%\newblock {\it Mon.\ Not.\ Roy.\ Ast.\ Soc.} {\bf 322}, 231--246 (2001)


\end{thebibliography}
\end{document}